\documentclass[sigconf]{acmart}
\acmConference[EASE 2025]{The 29th International Conference on Evaluation and Assessment in Software Engineering}{17–20 June, 2025}{Istanbul, Türkiye}
\AtBeginDocument{%
  }

\setcopyright{acmlicensed}
\copyrightyear{2025}
\acmYear{2025}
\acmDOI{XXXXXXX.XXXXXXX}
\acmISBN{978-1-4503-XXXX-X/2018/06}

\usepackage{hyperref} % For hyperlinks
\usepackage{xcolor} % For colors
\usepackage{tabularx}
\usepackage{graphicx}
\usepackage{pifont} 
\usepackage{tcolorbox} 
\begin{document}

\title{BugsRepo: A Comprehensive Curated Dataset of Bug Reports, Comments and Contributors Information from Bugzilla}

\author{Jagrit Acharya}
\orcid{0009-0008-0302-6130}
\affiliation{%
  \institution{University of Calgary}
  \city{Calgary}
  \country{Canada}
}
\email{jagrit.acharya1@ucalgary.ca}

\author{Gouri Ginde}
\orcid{0000-0001-7519-3503}
\affiliation{%
  \institution{University of Calgary}
  \city{Calgary}
  \country{Canada}
}
\email{gouri.deshpande@ucalgary.ca}


\begin{abstract}
Bug reports help software development teams enhance software quality, yet their utility is often compromised by unclear or incomplete information. This issue not only hinders developers' ability to quickly understand and resolve bugs but also poses significant challenges for various software maintenance prediction systems, such as bug triaging, severity prediction, and bug report summarization. To address this issue, we introduce \textnormal{{\fontfamily{ppl}\selectfont BugsRepo}}, a multifaceted dataset derived from Mozilla projects that offers three key components to support a wide range of software maintenance tasks. 
First, it includes a Bug report meta-data \& Comments dataset with detailed records for 119,585 fixed or closed and resolved bug reports, capturing fields like severity, creation time, status, and resolution to provide rich contextual insights. Second, {\fontfamily{ppl}\selectfont BugsRepo} features a contributor information dataset comprising 19,351 Mozilla community members, enriched with metadata on user roles, activity history, and contribution metrics such as the number of bugs filed, comments made, and patches reviewed, thus offering valuable information for tasks like developer recommendation. This contributor data further enables future research into expert identification within specific software modules, analysis of collaboration networks, prediction of bug resolution times based on assignee profiles, and studies on the impact of contributor experience on bug report quality. Lastly, the dataset provides a structured bug report subset of 10,351 well-structured bug reports, complete with steps to reproduce, actual behavior, and expected behavior. After this initial filter, a secondary filtering layer is applied using the CTQRS scale with additional filtering based on the CTQRS scale, a framework that uses dependency parsing and rule-based indicators to automatically assess the quality of the bug report by evaluating their morphological, relational, and analytical properties. By integrating static metadata, contributor statistics, and detailed comment threads, {\fontfamily{ppl}\selectfont BugsRepo} presents a holistic view of each bug's history, supporting advancements in automated bug report analysis, which can enhance the efficiency and effectiveness of software maintenance processes.

\end{abstract}




\maketitle

\section{Introduction}

Software maintenance is a crucial stage in software development, with bug resolution being one of its most significant tasks \cite{IT1}. Bug reports document the issues users experience while interacting with deployed software systems \cite{IT2}. However, the effectiveness of bug resolution techniques can be compromised when bug reports are incomplete or when automated systems rely on limited information from these reports. Research has shown that utilizing multiple aspects of bug reports for this task significantly improves the performance of predictive models and automated tools used in software maintenance \cite{Paikari2023,wang2025empiricalstudyleveragingimages}.

Bug reports are a fundamental component of bug resolution, but their effectiveness is frequently undermined by vague, incomplete, or ambiguous information provided by reporters \cite{fazzini2018automatically,good_report}. Clear and well-structured reports enable developers to quickly grasp the issue without extensive back-and-forth communication and enhance the quality of software maintenance tools\cite{wang2015}. Ideally, well-structured bug reports (also referred to as good/structured reports) should include the actual (incorrect) behavior (AB), the steps to reproduce the bug (S2R), the expected correct behavior (EB), and relevant environmental details, such as the software version and hardware specifications, as outlined in Bugzilla and GitHub bug reporting guidelines \cite{bugzilla_guidelines} \cite{githubWriteGood}. Missing critical information in incomplete reports often results in non-reproducible \cite{erfani2014works} and unresolved bugs \cite{zimmermann2012characterizing}, leading to delays in bug resolution and negatively impacting key software maintenance processes, including bug triaging, priority, severity prediction, and automated bug report summarization.

To address these challenges, we introduce  {\fontfamily{ppl}\selectfont BugsRepo}, a multifaceted dataset derived from Mozilla’s Bugzilla ecosystem. BugsRepo uniquely combines three components: (1) 119,585 curated bug reports with metadata \& comments (severity, resolution, lifecycle updates), (2) contributor profiles for 19,351 community members (tracking activity, patches, and collaboration patterns), and (3) a high-quality subset of 10,351 structured reports filtered using the regular expression \& CTQRS framework \cite{ctqrs}. CTQRS evaluates and filters reports based on morphological, relational, and analytical properties (e.g., clarity, completeness, and reproducibility). 
This contributor profile in particular, provides detailed metrics on user roles, activity history, and contributions (e.g., bugs filed, comments made, patches reviewed), enabling future works to build models for developer recommendation by matching bugs to developers with relevant expertise, predict bug resolution times based on contributor activity patterns, or analyze collaboration networks to optimize team dynamics. A scoring model can be introduced to provide a relevant score to each contributor, which will help in determining the severity and priority of the bug filed or comment made by the contributor. This integration of metadata, contributor context, and rigorously filtered reports creates a holistic resource for advancing software maintenance tasks.

Our work makes three key contributions:
\begin{itemize}
    \item \textbf{A scalable bug metadata \& comments dataset:} Covers over 50 Mozilla projects, enriched with status updates and resolution timelines (Size: 4.3 G.B., Dated: October 2024 ).
    \item \textbf{A contributor information dataset:} Captures roles, activity metrics, and collaboration networks to enhance developer recommendation systems.
    \item \textbf{A well-structured, high-Quality bug report subset:} Standardizes steps to reproduce (S2R), actual/expected behaviors, and environmental details, enabling robust model training for tasks like summarization and triage.
\end{itemize}

\begin{table*}[!htpb]
\centering
\renewcommand{\arraystretch}{1.2}
\caption{Overview of the Previous Work Dataset}
\label{Table:related_work}
\begin{tabular}{p{1.8cm} p{12cm} p{2cm}}
\textbf{Study} & \textbf{Description} & \textbf{Source}  \\ 
\hline
Kouzari \cite{Kouzari} & 
Event\_Id, Bug\_Id, Bug\_Description, User\_Email, 
Action\_datetime, Action\_Data, Priority, Resolution, Severity, Status & 
Bugzilla \\ 
\hline
Lamkanfi \cite{Lamkanfi} & priority, severity, product, component, bug status, resolution, assigned to, short desc, cc, version, op sys & Eclipse, Mozilla \\ 
\hline
Ahsan \cite{ahsan2010mining} & Developer Name/ID, Bug ID, Bug Title, Bug Severity Level, Bug Priority Level, Bug Assigned Date, Bug Resolved Date, Bug Fix Duration (Days), Total Working Days in a Month, Estimated Effort (Ei), Severity Weight (SW) & Eclipse, Mozilla, Linux, ArguUML and PostgreSQL  \\
\hline
Jiang et. al \cite{Authorship} & Developer Comments, Bug description, time stamp & Eclipse, Mozilla \\ \hline
Zhu et. al \cite{Multi} & Issue ,creation time, bug id, reporter, activity id, who, when, attribute, old value,new value 
Comment\_id,who,when,text  & Eclipse, Mozilla \\ \hline
\textbf{Our dataset} & \textbf{Bug metadata:} Bug ID, Summary, Priority, Severity .. \& \textbf{Contributor Information Dataset}: Bug ID, Comment ID, Author, Comment Text, Author ID, User Name, Created On, Last Activity, Permissions, Bugs Filed, Comments Made, Assigned to, Assigned to and Fixed, Commented on, QA Contact, Patches Submitted, Patches Reviewed, Bugs Poked \& \textbf{Comments} \& \textbf{Discussion Dataset}: Bug ID, Comment ID, Author, Comment Text, Bug Report & Bugzilla  \\
\end{tabular}
\end{table*}

\section{Novelty of BugsRepo }

A key novelty of {\fontfamily{ppl}\selectfont BugsRepo} is its holistic integration of rigorously curated data, combining bug report meta-data, discussion comments, structured bug reports and contributor information. This multi-faceted design addresses longstanding challenges in software maintenance by providing researchers and practitioners with a comprehensive foundation for tasks such as bug summarization, severity prediction, and developer recommendation. We apply dual filtering of bug reports to retain only those containing well-written S2R, AR, and ER sections, thereby removing the reports that lack the important information needed for the various bug report management tasks. Our approach leverages regular expressions to detect the presence of these fields. Additionally, we integrate CTQRS \cite{ctqrs}, which uses dependency parsing and rule induction to automatically assess the quality of bug reports. CTQRS explores key desirable properties of bug reports, including atomicity, correctness, completeness, conciseness, understandability, and reproducibility. This ensures that the dataset consists solely of high-quality, structured reports, which is crucial for software maintenance tasks such as bug triaging and severity prediction etc. As a result, the curated subset of 10,351 reports can be regarded as a higher-quality corpus compared to typical Bugzilla data. Such structured bug reports allow researchers to train and evaluate ML models (for triage, summarization, severity prediction, etc.) on more reliable data, reducing noisy results caused by partial or incomplete bug descriptions. Further, to enhance the bug reports data, we have collected 19,351 Mozilla community members, with details on their contribution toward the organization (i.e Bug resolved, Comments made etc), which may help in assigning bug reports, priority, and severity of the tasks.

\section{Data Description}
Our dataset comprises of an extensive collection of bug report-related data accumulated over the past five years. It includes comprehensive bug metadata covering various aspects such as severity, status, priority, product version, platform, classification, etc. Beyond static metadata we introduced fields like the ``contributor\_Id \& ``contributor\_email" to record all the contributors related to a particular bug. The dataset captures all lifecycle status updates and detailed comment threads that document the discussions and content exchanged throughout the bug resolution process. Additionally, it features a contributor information dataset for both commenters and bug reporters, providing details regarding user activities like the number of bugs filed, comments made, and patches submitted. The Comments dataset provides a complete view of comments and discussion on each bug and its history, enabling thorough analysis and promoting a better understanding of the bug-tracking system within the Mozilla project. To further enhance software maintenance tasks, we also provide a subset dataset of high-quality, well-structured bug reports dataset with complete Steps to reproduce, actual behavior, expected behavior and additional information for advancing in software maintenance processes.

By offering such detailed insights over meta-data, contributor information, comments and well-structured reports, the group of datasets may enhance the possibilities for advanced bug report analysis in bug triaging, root cause identification, priority prediction, severity prediction, duplicate prediction, fixing time estimation, categorization, summarization, impact analysis and bug report life cycle analysis domains.

BugsRepo distinguishes itself from existing datasets summarized in Table \ref{Table:related_work}, which often focus primarily on bug metadata or comment text \cite{Kouzari, Jiang}, by providing a unique, integrated collection that includes comprehensive bug metadata, the full history of discussion comments of that bug, and a rich Contributor Information Dataset detailing user roles, activity patterns, and contribution metrics for over 19,000 community members; this holistic approach allows researchers to move beyond isolated analyses and explore the complex interplay between bug characteristics, communication dynamics, and the expertise and historical contributions of individuals, which is essential for developing more intelligent and context-aware software maintenance tools like advanced bug triaging and personalized developer recommendation systems.

The dataset (4.3 GB) can be obtained from the following link\footnote{\url{https://zenodo.org/records/15004067}}.
The code to mine our dataset and rate the quality of bug reports can be obtained from the following link\footnote{\url{https://github.com/GindeLab/EASE_2025_Data_paper}}.

\begin{figure}[!t]
    \centering
    \includegraphics[width=\linewidth]{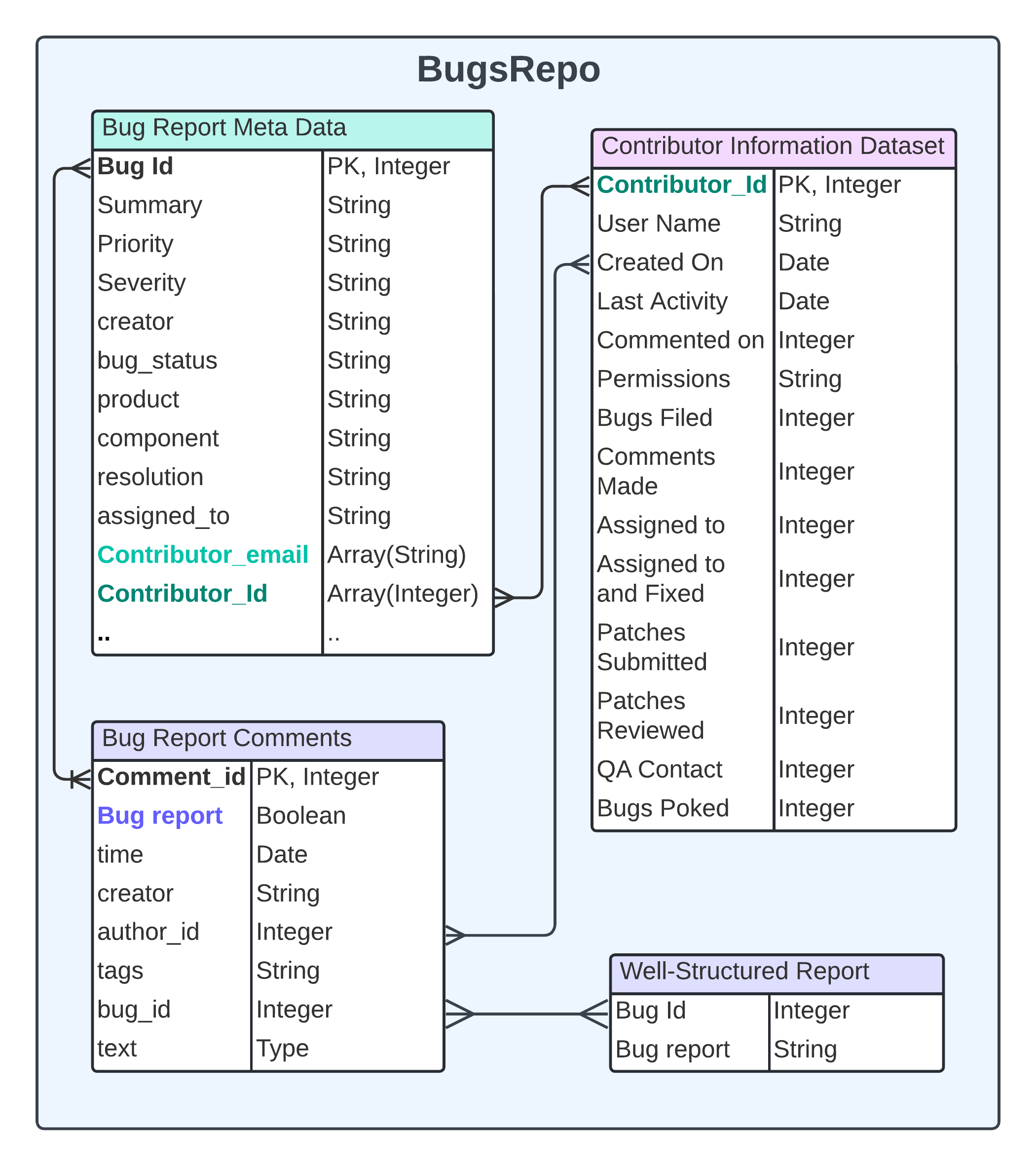} % Replace "example-"mage" with th" path to your image file
    \caption{
    Entity-Relationship Diagram depicting a many-to-many relationship among bug Report metadata, bug report comments, and contributor information dataset, illustrating columns and data types within our analyzed dataset.    }
    \label{fig:ER_diagram}
\end{figure}
% \section{Contribution}
% This paper makes three main contributions:
% A bug report metadata dataset of Mozilla Products called \textbf{BugsRepo} consisting of 50,746 bugs from popular projects.
% A Contributor Reputation dataset consisting of 20,124 members of Mozilla community,

% A Bug Report dataset of 30,122 reports of well-structured bugs with Steps to produce, Actual behaviour, expected behaviour and Additional information.
Our contributions are as follows, collectively named {\fontfamily{ppl}\selectfont BugsRepo}: 
 
\begin{itemize}
    \item A \textbf{Bug report meta-data \& comment Dataset}, which includes detailed records of 119,585 bugs from more than 50 Mozilla Projects based on all data gathered as of October 2024. Each record captures comprehensive fields such as severity, creation time, status, and resolution, enriching the context for each bug report.
    \item A \textbf{Contributor Information Dataset} consisting of 19,351 Mozilla community members, which contains details on user roles, activity history, and contributions metrics such as number of bugs filed, comments made, and patches reviewed, providing a deep dive into community engagement and expertise.
    \item A \textbf{Structured Bug Report Dataset and Comments} comprising 10,351 bug reports. This dataset of well-structured reports includes steps to reproduce, actual behavior, expected behavior, and additional information, along with discussion comments offering a well-rounded view of each bug's impact and nature.
\end{itemize}
\begin{figure}[h]
  \centering  \includegraphics[scale=.35]{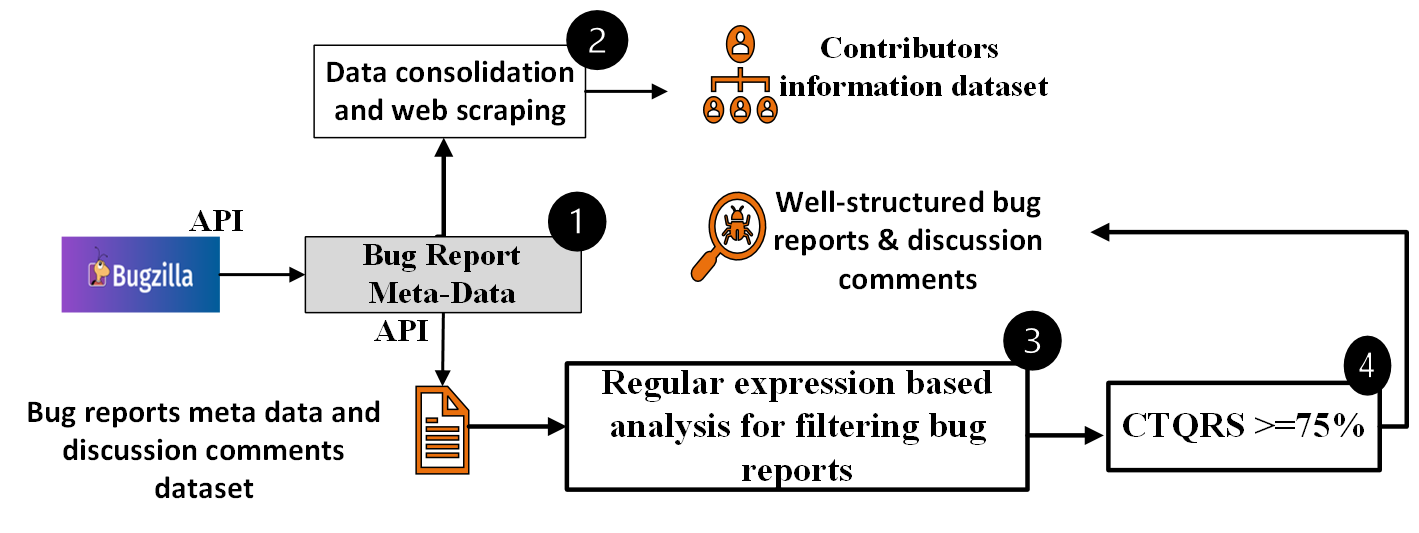}  % Adjust the image file name and path as needed
  \caption{ Overview of the methodology used to develop various datasets}
  \label{fig:Meth}
  \vspace{-4mm}
\end{figure}
% \begin{figure}[h]
%   \centering  \includegraphics[scale=.8]{Images/Good_reports.png}  % Adjust the image file name and path as needed
%   \caption{ Well Structured Bug reports VS projects distribution, showing Core and Firefox are projects with most well-structured bug reports filed in last 5 years}
%   \label{fig:good_reports}
% \end{figure}

\textbf{CTQRS:} We implemented the CTQRS (Crowdsourced Test Report Quality Score) framework using the Stanza library to assess bug report quality. CTQRS, developed by Zhang et al. \cite{ctqrs}, evaluates reports based on three key indicators: morphological (like size, readability, and punctuation), relational (such as itemization, environment details, and screenshots), and analytical (which focus on interface elements, user behavior, and defect descriptions). These indicators help evaluate properties like completeness, clarity, and reproducibility, with a maximum score of 17 points. To achieve this, we used Stanza for natural language processing tasks, including word segmentation, part-of-speech tagging, syntactic and semantic dependency parsing, and semantic role labeling.
These techniques enabled us to analyze the structure and meaning of bug reports effectively, ensuring a more accurate assessment of their quality. The authors set a 75\% threshold as a benchmark for high-quality, clear, and complete reports. Accordingly, we have adopted this threshold as our benchmark.

% These techniques allowed us to analyze the structure and meaning of bug reports effectively, ensuring a more accurate assessment of their quality. The author analyzed reports that scored above 75\% with the 75\% threshold set by the authors as a benchmark for high-quality, clear, and complete reports.. Therefore, we have used this threshold as our benchmark.
% CTQRS good reports number 10,351

\section{Data collection} 
\begin{figure}[h]
  \centering
  \includegraphics[scale=.3]{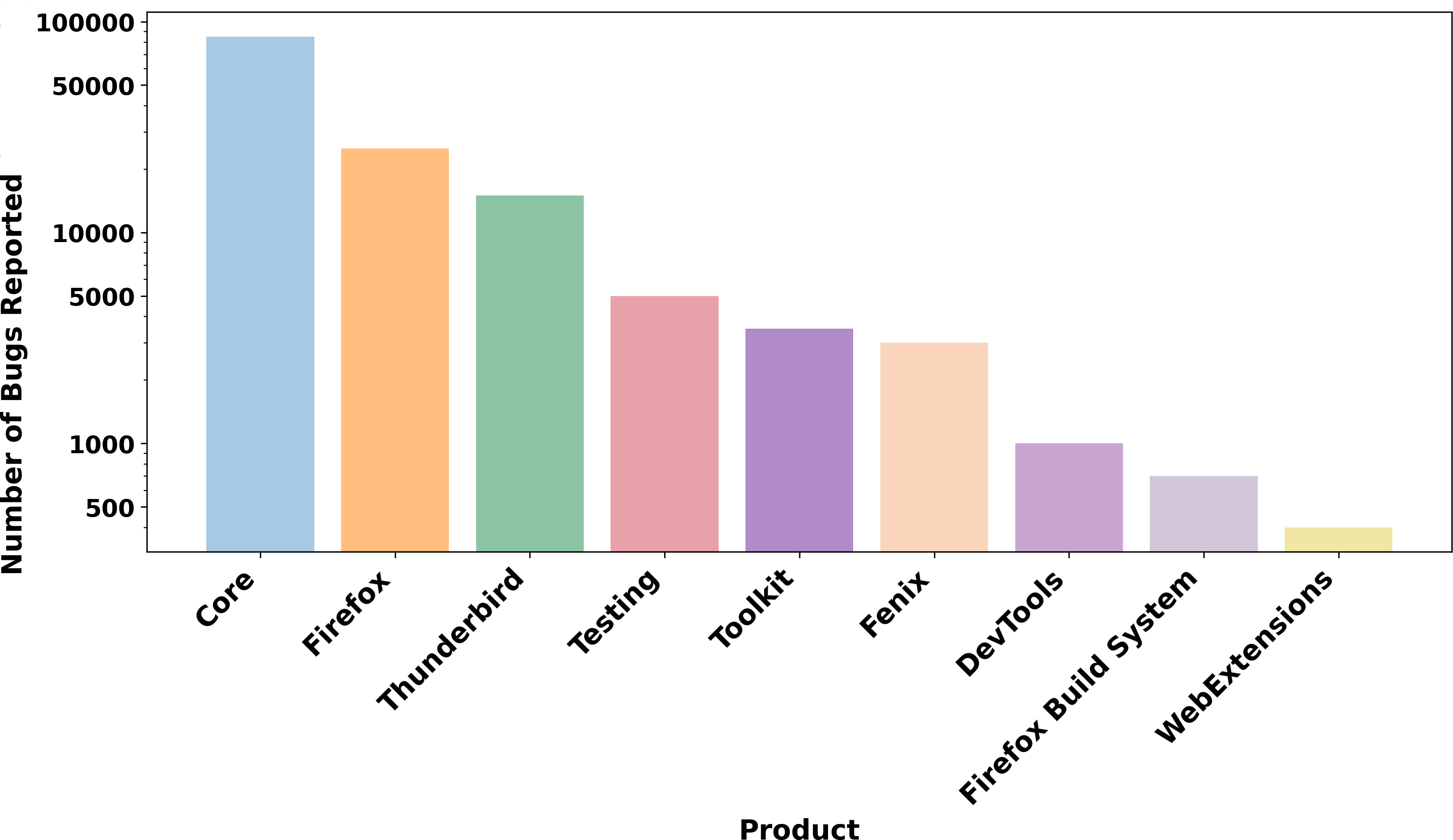}
  \vspace{-10pt}  % Reduce vertical space after image
  \caption{Bug reports vs. projects distribution, showing Core and Firefox are projects with most bug reports filed in last 5 years}
  \label{fig:Analysis}
\end{figure}

This section outlines the methodology employed to develop {\fontfamily{ppl}\selectfont BugsRepo} dataset. Figure \ref{fig:Meth} illustrates an overview of the methodology workflow, which is divided into four primary phases: collecting bug reports metadata, gathering the contributor information dataset, collecting the actual bug reports and discussion comments used for summarization of whole discussion, and filtering the well-structured reports. The following subsections provide a detailed explanation of each phase. The product distribution of our raw dataset is shown in Figure \ref{fig:Analysis}, highlighting Core, Firefox and Thunderbird as the top three projects with the most number of bugs getting reported. The detailed use case of each project is shown in Table \ref{Table:MozillaProjects}, further details about the projects are available at GitHub\footnote{\url{https://github.com/GindeLab/EASE_2025_Data_paper}}.

\begin{table}[!htpb]
  \centering
  \renewcommand{\arraystretch}{1.1}
  \caption{Overview of Key Mozilla Projects}
  \begin{tabular}{p{2cm} p{6cm}}
    \textbf{Project} & \textbf{What it does} \\ \hline
    Core & Show websites and run JavaScript in Firefox using the Gecko engine and SpiderMonkey \\ \hline
    Firefox & A web browser focused on speed, privacy, and user control \\ \hline
    Thunderbird & Free email app with calendar and contact tools \\ \hline
    Mozilla Testing & Runs checks on Mozilla software to find bugs early \\ \hline
    Mozilla Toolkit & Tools for building apps that run on any system \\ \hline
    Fenix \newline (Firefox Android) & New Firefox for Android using GeckoView engine \\ \hline
    DevTools & Tools inside Firefox for editing and debugging websites \\ \hline
    Firefox \newline Build System & System to turn Firefox source code into apps \\ \hline
    Web Extensions & One system to build browser add-ons for multiple browsers \\ \hline
  \end{tabular}
  \label{Table:MozillaProjects}
\end{table}

\noindent\textit{Info on additional projects is available at \url{https://github.com/GindeLab/EASE_2025_Data_paper}.}\\

In the bug reports collection phase, we built a dataset of bug reports based on Bugzilla \cite{bugzilla}. We selected Bugzilla as the source for building {\fontfamily{ppl}\selectfont BugsRepo}, both due to its widespread use and dedicated focus on issue tracking, which allowed us to mine relevant bug reports effectively \cite{kumarcomparative}. To identify relevant issues, we developed a Python script utilizing the Bugzilla REST API \cite{mozillaBugzillaREST2024} to extract entries containing reported bugs as shown in \ding{202} of Figure \ref{fig:Meth}.

After retrieving the bug report metadata, we created our web scrapping script using Beautiful Soup \cite{crummyBeautifulSoup} to fetch contributor information data listed on the Bugzilla website, depicted in \ding{203} of Figure \ref{fig:Meth}. As previously explored by many researchers \cite{Web_s_2} and Ardimento et al. \cite{Web_scrap1}, using the R software system \cite{rprojectProjectStatistical} has proven effective for similar web scraping tasks, enabling efficient data extraction and analysis.

We then leveraged the Bugzilla API \cite{mozillaBugzillaREST2024}to collect discussion comments and actual bug reports, as illustrated in Figure \ref{fig:Meth}. However many of the filed bug reports are unclear or incomplete details weaken their usefulness. Wang et al. shows that good quality, clear, well-structured reports help developers understand issues faster \cite{wang2015}. A good report should include:  
\begin{itemize}  
    \item \textbf{Steps to Reproduce (S2R)}: How to recreate the bug.  
    \item \textbf{Actual behavior (AR)}: What is currently wrong.  
    \item \textbf{Expected behavior (ER)}: What should happen instead.  
\end{itemize}  
Environmental details (e.g., software version) are also critical, as noted in guidelines from Bugzilla and GitHub \cite{Mozilla_Bugzilla_2024, githubWriteGood}. Poorly written as explained in Figure \ref{red_input}, S2Rs make bugs hard to replicate, delaying fixes \cite{fazzini2018automatically, good_report}. Missing information leads to non-reproducible \cite{erfani2014works} and unresolved bugs \cite{zimmermann2012characterizing}.
Hence to make filter-incomplete bug reports, we used regular expressions.
Our analysis of 119,585 raw bug reports from Bugzilla found that only 12,614 fully complied with Bugzilla's reporting guidelines \cite{bugzilla_guidelines}, which advises reporters to include steps to reproduce (S2Rs), actual behavior, expected behavior, and additional relevant information in their bug report. Thus, using regular expressions, we filtered reports including steps to reproduce, actual and expected behavior, focusing on bugs with `RESOLVED' or `CLOSED' status and `FIXED' resolution, using a regular expression as shown in \ding{204} of Figure  \ref{fig:Meth}.
To further improve dataset quality, we applied filtering based on the CTQRS score \cite{ctqrs}, as demonstrated in \ding{205} of Figure \ref{fig:Meth}, reducing the dataset to 10,351 reports. 
Bug report comments are essentially important as they help in determining the priority and severity based on context and criticality of the discussion. It can help summarizing the whole discussion between the reporters and the developers.

\begin{figure}[h!]
    \centering
    \begin{tcolorbox}[colback=green!5!white, colframe=green!75!black, title=Sample well-structured bug report]
 \textbf{Bug ID: 1889116}\newline
    \textbf{Steps to Reproduce:}\newline
    1. On the home screen, tap the search engine icon drop-down menu.\newline
    2. Select Tabs/Bookmarks/History search engine.\newline
    3. Observe the selected search icon from the search bar.\newline
    4. Select Bing/Amazon/DuckDuckGo/eBay/Wikipedia engine.\newline
    5. Observe the selected search icon from the search bar.\newline
    
    \textbf{Expected Behavior:}\newline
    The search icon is successfully updated according to the selected search engine.\newline
    
    \textbf{Actual Behavior:}\newline
    a) There is no icon displayed when selecting one of the Tabs/Bookmarks/History options. (Reproducible only in dark mode)\newline
    b) The first selected search icon remains displayed for all the next selected engines.\newline
    
    \textbf{Additional Information:}\newline
    * Firefox Version: Firefox Nightly 126 (2024-04-02)\newline
    * Android Device Model: Samsung A32 (Android 13), Oppo A15s (Android 10), Xiaomi Mi8 Lite (Android 10)\newline
    * This issue is a recent regression: Last good build (2024-04-01) - First bad build (2024-04-02).\newline
    \end{tcolorbox}
    \caption{This is an example of a high-quality, well-structured bug report. The report contains complete steps to reproduce, expected behavior, actual behavior and additional information.}
    \label{green_output}
       \vspace{-3mm}
\end{figure}
\begin{figure}[h!]
    \centering
    \begin{tcolorbox}[colback=red!5!white, colframe=red!75!black, title=Sample unstructured (lacking standard format) bug report]
    \textbf{Bug Id: 1805934} 

User Agent: Mozilla/5.0 (X11; Linux x86\_64; rv:108.0) Gecko/20100101 Firefox/108.0

Steps to reproduce:

For a few months now, I've been suffering an intermittent problem: every now and again, all drop-down controls in Firefox would break. Menus would no longer work, drop-down selects on web pages would fail, extension menus would fail, and the hamburger menu would fail. The visible behavior is that the drop-down is drawn but then immediately erased as if I had clicked elsewhere in the window. The only fix for the problem is to restart Firefox.

Recently, I realized something: every time I restarted to fix the issue, Firefox would bring up the dialog saying it was installing the latest update. And I never get a dialog to tell me that an update is available.

So what seems to be happening is: every time Firefox detects an available update, something breaks and all menus and drop-downs stop working.

Today was even worse: restarting didn't show the updating dialog, and as soon as I went to any web page, all the drop-downs broke again. So I wondered if I was wrong about the cause... but I cleared all local data (cache, cookies, the lot) and restarted one more time -- and suddenly I got the updating dialog, and now drop-downs work again.

Obviously this is absolutely infuriating. I'd like to do anything I can to help you track down and fix the problem.
   \end{tcolorbox}
\caption{\textbf{This is an example of a low-quality bug report, as it does not follow the defined Bugzilla bug report template.}}

    \label{red_input}
    \vspace{-3mm}
\end{figure}
% \section{MOTIVATION}
% bugzilla is one of the most favoured data sources among researchers studying bug report processing techniques, with many utilizing its datasets from prominent projects like Eclipse, Mozilla, and Open Office to evaluate their proposed methods.\cite{Lee2019BugReport}

% Bug reports are crucial in software development, but their effectiveness is sometimes reduced by the lack of clear, complete, or specific information provided by the reporters. When reports are well-structured and clear, developers can quickly grasp the problem without needing extensive discussions \cite{wang2015}.
% For instance, effective bug reports should detail the observed incorrect behaviour (OB), the steps needed to reproduce the bug (S2R), the expected correct behaviour (EB), and important details like software version and hardware specifications, following the guidelines from bugzilla and GitHub \cite{bugzilla_guidelines} \cite{githubWriteGood}.
% However, a lot of reports have unclear, incomplete, or vague reproduction steps, which makes it difficult for developers to duplicate and fix the issues in the software \cite{fazzini2018automatically,good_report}.

\begin{table}[!htpb]
  \centering
  \renewcommand{\arraystretch}{1.1}
  \caption{Metadata for a bug report}
  \begin{tabular}{p{2cm} p{4cm} p{1cm}}
    \textbf{Property} & \textbf{Description} & \textbf{Data type} \\ \hline
    \textit{Bug ID} & Unique identifier of a bug report & Number \\ \hline
    Summary & A brief title of the bug report & String \\ \hline
    Product & The product or project the bug/issue is associated with & String \\ \hline
    Component & The component of the product the bug is related to & String \\ \hline
    Version & The version of the product affected & String \\ \hline
    Priority \& Severity & The priority \& severity level of the bug & String \\ \hline
    .. & .. & .. \\ \hline
  \end{tabular}
  \label{Table:Datacontent}
\end{table}

\vspace{1cm} % Increase vertical spacing between tables

\begin{table}[!htpb]
  \centering
  \caption{Detailed table of contributor information dataset}
  \begin{tabular}{p{1.5cm} p{4.8cm} p{1.2cm}}
    \textbf{Field} & \textbf{Description} & \textbf{Data type} \\ \hline
    \textit{Bug ID} & Unique identifier for the bug associated with the comment & Integer \\ \hline
    Comment text & Text content of the comment & Text \\ \hline
    User name & User name of the comment author & String \\ \hline
    Last activity & Date and time of the last activity on the comment & DateTime \\ \hline
    Bugs filed & Number of bugs filed by the author & Integer \\ \hline
    Assigned to and fixed & Number of bugs assigned to the author that have been fixed & Integer \\ \hline
    .. & .. & .. \\ \hline
  \end{tabular}
  \label{tab:my_label}
\end{table}

\begin{table}[!htpb]
  \centering
  \renewcommand{\arraystretch}{1.1}
  \caption{Overview of the basic comment dataset}
  \label{Table:BasicCommentData}
  \begin{tabular}{p{1.7cm} p{4.5cm} p{1.3cm}}
    \textbf{Column} & \textbf{Description} & \textbf{Data Type} \\ \hline
    Bug ID & Unique identifier for the bug associated with the comment & Integer \\ \hline
    Comment ID & Unique identifier for the comment within the bug report & Integer \\ \hline
    Author & Name or identifier of the person who made the comment & String \\ \hline
    Comment text & The initial comment contains the bug report, and subsequent comments have the discussion & Text \\ \hline
    Bug report & It comment text contains the bug report & Boolean \\ \hline
  \end{tabular}
\end{table}
% \vspace{15pt}
\section{Applications}
Our dataset is primarily designed to support research focused on automating the analysis, comprehension, and generation of bug reports. Bug report management tasks are used at different stages of the bug report lifecycle \cite{Zhang2015}. Starting with Bug Triage and Bug priority-severity prediction, followed by bug classification and summarization.

\textbf{Bug Triage:} Bug Triage refers to assigning a bug to a developer and is widely explored among researchers, using Graphs \cite{DAI2023111667}, SVM-based approaches \cite{Anvik}, deep learning methods (DBRNN-A \cite{Mani}), graph-based models (DABT \cite{Jahanshahi}, PMI \cite{Zaidi}), and GRCNN \cite{WU2022108308}). Using bug title, description, developer email, reported time/status, filtered by fixed/verified statuses for Google Chromium, Mozilla Core, and Mozilla Firefox as their datasets.
We believe BugRepo can enhance triage in four ways 
\begin{itemize}
    \item Structured Bug Context: S2R, ER, and AR sections reduce noise in training data, improving model accuracy in predicting developer.
    \item Contributor Expertise: Developer activity metrics (e.g., Activity by product, patches reviewed, bugs fixed etc) enable skill-based triage, matching bugs to developers with domain-specific expertise.
    \item Workload Balancing: Contributor metadata (e.g., ``Bugs Assigned" and ``last activity") enables workload-aware triage to prevent developer burnout.
    \item Cross-Project Triage: With data from 50+ Mozilla projects, models can generalize triage policies across multiple ecosystems of projects.
\end{itemize}

% \textbf{Bug Duplicate Detection:} The goal is to detect if the bug report has already been filed by any reporter. Multiple Machine Learning \cite{ml_Budhiraja2018,ml_Hu2018,Ml_Gopalan2014}, Deep Learning techniques \cite{ZHOU2020110572,ZHENG2024111258}. Most of the researchers are using similar datasets comprising of  product, component, platform, status, priority, summary, and description from encompassing spark, eclipse, mozilla firefox, thunderbird, and other Mozilla projects. This can be enhanced with a well-structured bug report dataset that includes detailed comment analyses—featuring timestamps, privacy statuses, attachment data, and the raw text of communications—along with contributor metadata detailing each user's bug reporting history and activity. Additionally, if researchers incorporate comprehensive bug attributes such as severity, status, resolution, and relevant timestamps along with author information, it may expand the research's analytical depth and insights.
% allowing for a more nuanced feature set that can significantly improve the efficiency of duplicate detection. 

\textbf{Bug Report Summarization:} Bug report summarization involves condensing the content of bug reports and their associated comments into concise summaries, highlighting essential information to facilitate quick understanding and decision-making. 

Bug report comments typically appear after a bug is filed, so early-stage triage or severity assignment might not benefit from them immediately. However, in many real-world processes, bug triage can be a multi-step procedure. As more discussion emerges, advanced tasks (like severity re-evaluation or updated triage) become possible. Consequently, the availability of comment history in {\fontfamily{ppl}\selectfont BugsRepo} remains valuable for iterative or incremental approaches to bug management. The comments in our dataset can also used in summarizing the discussion over the bug report.
SumLLaMA \cite{xiang2024sumllama} leverages contrastive learning pre-training and efficient fine-tuning of large language models (LLMs) to enhance summarization performance. Few advanced methodologies such as DeepSum \cite{li2018unsupervised}, which utilizes neural networks and word2vec embeddings; BugSum \cite{liu2020bugsum}, which integrates Bi-directional Gated Recurrent Units (Bi-GRUs). All these approaches utilize textual elements like titles, product details, components, descriptions, and authorship data to generate summaries. We believe BugRepo advances in this via:

\begin{itemize}
    \item Comment Weighting: Contributor ``score" (from bugs fixed,) prioritizes expert comments in summaries.
    \item Structured Section Integration: Directly add S2R/ER/AR sections into summaries for reproducibility.
    \item Multi-Modal Summaries: Combine text with attachment metadata (e.g., product, component) along with contributor information for richer context of summaries.
\end{itemize}

\textbf{Bug Priority \& Severity Prediction:}
 In this problem domain, the focus is on determining the order in which bugs should be addressed (priority) and assessing the urgency of these issues (severity). Numerous Machine Learning \cite{pri_ml_ahmed2021capbug,pri_ml_huang2022bug,pri_ml_shatnawi2022assessment} and Deep Learning-based \cite{dl_pri_bani2021deep,dl_pri_fang2021effective} automatic software priority prediction techniques have been proposed in the literature using the textual information available such as bug report summary, product, component in software bug reports. We believe BugRepo dataset allows
 \begin{itemize}
    \item  Reporter Reputation: Contributor ``Bugs Resolved" metrics gauge reporter credibility, refining severity estimation.
    \item  Comment Sentiment: Analyze comment text (via metadata) to detect urgency using natural language processing.
    \item  Community Trust Networks: Model contributor relationships (e.g., frequent collaborators) to identify consensus on priority of the bug report.
\end{itemize}

 \section{Challenges and Limitations}
 \textbf{Challenges:} One of the initial challenges we faced in this study involved the creation of the contributor information dataset. To compile this dataset, we developed our own web scraper using Beautiful Soup \cite{crummyBeautifulSoup}, a popular Python library for parsing HTML and XML documents. The development of the web scraper required multiple iterations to accurately navigate through the web page and extract the necessary data and dealing with periodic changes in page structure was challenging. Identifying the correct HTML tags, such as classes and IDs, proved to be a complex task due to the dynamic nature of web content. Each iteration involved refining our scraping logic to adapt to these inconsistencies and ensure the reliable extraction of data.  Additionally, we faced difficulties with Bugzilla API and web pages that would block our requests if too many were made in a short period, treating our scraper as a potential bot. To mitigate this, we implemented rate-limiting measures, including a sleep command in our script, ensuring that no more than 60 requests were made per minute.\\
 
\textbf{Limitations:} The dataset comprises bug reports submitted from 2018 onwards, covering reports up to October 2024 for Mozilla. It is important to note that Bugzilla allows users to request new features, which are not considered actual bugs. To maintain focus on genuine bug reports, entries with the severity attribute marked as ``enhancement" (indicating feature requests) are excluded. Additionally, Bugzilla tracks whether reports are resolved or unresolved. Only resolved reports are included in this dataset, as they represent bugs that have completed their entire life cycle. Over 1 million bugs have been reported on Bugzilla, making it impractical to manually gather and analyze all this data.
\section{Related Work}
Bugzilla is one of the most favoured data sources among researchers studying bug report processing techniques, with many utilizing its datasets from prominent projects like Eclipse, Mozilla, and Open Office to evaluate their proposed methods \cite{Lee2019BugReport}. Previous research has made significant progress in sharing data extracted from issue tracking systems, utilizing datasets such as the Eclipse and Mozilla defect tracking datasets capturing metadata \cite{Lamkanfi}, the Firefox temporal defect dataset \cite{habayeb2015firefox} capturing the essential temporal events throughout a bug's lifecycle, eliminating the need to mine its history files. Zhu et al. \cite{Multi} mined a 15-year Mozilla Bugzilla dataset featuring multiple extracts and data levels attributes discussed in Table \ref{Table:related_work}; these datasets are meticulously curated to address specific research questions, involving processes like filtering, cleaning, and transforming the data to enhance usability for researchers.
% We saw that there is a lack of a bug reports dataset specific to bugzilla, which follows their guidelines. We introduce 19,351 contributors to bugzilla's information dataset, which can significantly impact the research in bug report analysis.  

However, we observed that there is a lack of a bug reports dataset specific to Bugzilla that strictly follows their guidelines. We also introduce a dataset of 19,351 contributors to Bugzilla's information, which can impact research in bug report analysis.

Bug report quality has been widely studied, with various methods proposed to evaluate and improve it \cite{good_report,hao2019ctras}. Many of these methods use heuristic rules and expert insights to identify key details in bug reports. He et al. \cite{he2020deep} introduced a convolutional neural network (CNN)-based approach to classify bug reports as valid or invalid using only textual data, such as summaries and descriptions. Similarly, Chen et al. \cite{chen2018automated} leveraged natural language processing and quantifiable indicators to assess bug report quality automatically. Building on this, we adopted the CTQRS framework by Zhang et al. \cite{ctqrs}, which uses dependency parsing and rule-based indicators to automatically assess the quality of crowdsourced test reports by evaluating their morphological, relational, and analytical properties. It examines factors such as size, readability, punctuation, interface element descriptions, user behavior, and system defect details to assign a quality score based on predefined rules and quantifiable indicators.

\subsection{Comparison with Existing Datasets}
Numerous datasets have emerged from Eclipse, Open Office, and Thunderbird bug repositories \cite{Lamkanfi,ahsan2010mining,Authorship}, these resources often focus on only high-level bug metadata (e.g., severity, priority, and summary) or lack comprehensive life-cycle updates. In contrast, {\fontfamily{ppl}\selectfont BugsRepo} uniquely combines (1) a large-scale set of Mozilla bug reports with fully detailed metadata (2) contributor information capturing user roles, comments made, bug filed, and submitted patches etc, and (3) well structured bug reports that strictly follow Bugzilla’s guidelines \cite{Bugzilla_guidelines_working} for Steps to Reproduce (S2R), Actual behavior (AB), and Expected behavior (EB). This three-pronged approach provides deeper insights into software maintenance, especially for tasks like developer recommendation and advanced summarization and many more.
\section{Conclusion}
In this paper, we present, {\fontfamily{ppl}\selectfont BugsRepo}
a comprehensive dataset mined from Mozilla projects, integrating three distinct datasets. We detail the mining process and the tools utilized for data extraction. Additionally, we provide a replication package containing the code used to mine the reports. Three datasets are presented, each tailored to different aspects of bug analysis: a bug report meta-data dataset, a contributor information dataset, and a discussion comments \& structured bug report dataset. We also discuss a characterizing analysis of how our datasets can be used to enhance bug report summarization, prioritize bug severity, and improve the accuracy of automated bug triage systems.

\bibliographystyle{ACM-Reference-Format}
\bibliography{sample-base}

% that's althat'sks
\end{document}